\documentclass[aps,prc,floats,floatfix,preprint,superscriptaddress,nofootinbib]{revtex4-1}  
\usepackage{graphicx}
\usepackage{epstopdf}
\usepackage{color}
\begin{document}

\def\be{\begin{equation}}   
\def\ee{\end{equation}}

\newcommand{\affA}{
Graduate School of Pure and Applied Sciences, University of Tsukuba, 
Tsukuba 305-8571,Japan
}
\newcommand{\affB}{
Center for Computational Sciences, University of Tsukuba, 
Tsukuba 305-8577, Japan
}

\title{Efficient basis expansion for describing linear and nonlinear electron 
dynamics in crystalline solids}

\author{Shunsuke A. Sato}\affiliation{\affA}
\author{Kazuhiro Yabana}\affiliation{\affA}\affiliation{\affB}

\begin{abstract}
We propose an efficient basis expansion for electron orbitals 
to describe real-time electron dynamics in crystalline solids.
Although a conventional grid representation in the three-dimensional 
Cartesian coordinates works robustly, it requires a large amount 
of computational resources. To reduce computational costs, we consider 
an expansion using basis functions with a truncation.
A simple choice employing eigenstates of the ground state 
Hamiltonian with a truncation turned out to be useless.
We have found that adding occupied eigenstates of nearby 
$k$-points to the truncated basis functions composed of 
eigenstates of the original $k$-point is crucially important. 
We demonstrate the usefulness of the method for linear and nonlinear 
electron dynamics calculations in crystalline SiO$_2$.

\end{abstract}
\maketitle
\section{Introduction}

In current frontiers of optical sciences, interactions of
intense and/or ultrashort laser pulses with crystalline solids
have been attracting substantial interests from both fundamental
and technological points of view 
\cite{ps99,bk00,cm07,ga11,mi11,ba13,sb13,sp13}.
To describe interactions of visible or infrared laser pulses 
with crystalline solids, it is necessary to calculate 
time evolution of electron orbitals in a unit cell of solids
under a time-varying, spatially-uniform electric field. 

In the first-principles level, calculations based on the
time-dependent density functional theory (TDDFT) \cite{ru84}
have been carried out.
We have been developing a computational method solving
the time-dependent Kohn-Sham (TDKS) equation in real time
in a unit cell of solids in which a uniform grid representation 
in the three-dimensional Cartesian coordinates is used to 
express orbital wave functions 
\cite{be00,ot08,sh10,sh12,sa14}.
A similar computational approach has also been developed and
utilized in Ref.\cite{wa11,wa12,wa13,go13}.
A computational method employing plane wave basis
has also been developed \cite{ta06}.
These computational schemes are applicable to electron
dynamics in linear regimes \cite{be00,ta06} and nonlinear electron
dynamics under intense laser pulses \cite{ot08,sh10,wa11,sh12,wa12,wa13,sa14}. 
There are also a number of works solving the time-dependent
Schr\"odinger equation in empirical periodic potentials
\cite{pi92,ba06,ap12,ko13}. 

Although numerical methods employing a uniform grid representation 
in the three-dimensional Cartesian coordinates works stably and 
robustly, time evolution calculations in this representation are 
computationally demanding. In particular, the required 
computational resource is excessively large when we couple the 
electron dynamics with the dynamics of macroscopic electromagnetic 
fields \cite{ya12,le14}.
We are thus urged to find a more efficient representation in
which computational costs are reduced without harming numerical accuracy.
In molecules and nano-sized materials, orbital expansions are often
utilized for real time evolution calculations. For example in 
Ref.\cite{go11}, time evolution calculations are achieved in a 
dense matrix representation employing truncated eigenfunctions 
of the static Kohn-Sham Hamiltonian with spatial-domain 
decompositions.

As basis functions in crystalline solids, one may suppose that 
a finite number of eigenstates of the static Kohn-Sham Hamiltonian 
at each $k$-point would serve as a natural choice.
However, as will be shown, this truncation scheme does not work well.
We will show that it is crucially important to use basis functions
in which occupied eigenstates of nearby $k$-points are added to the 
eigenstates of the original $k$-point.
We will illustrate usefulness of the method in the time evolution
calculation taking a system of SiO$_2$ described with
a time-independent periodic potential.

The construction of the paper is as follows.
In Sec. II, we describe theoretical frameworks and our proposal for 
the efficient basis functions. In Sec. III, we examine usefulness
of the proposed basis expansion for linear response problems.
In Sec. IV, applications to electron dynamics under pulsed electric
fields is discussed.
In Sec. V, we discuss computational aspects of the proposed method. 
In Sec. VI, a summary will be presented.

\section{Basis functions for electron dynamics calculations}
\subsection{Time-dependent Schr\"odinger equation}

We consider electron dynamics in a crystalline solid under an optical 
electric field of visible or infrared frequencies. 
We assume a long wavelength limit: The wave length of the optical electric 
field is much longer than the length scale of the electron dynamics 
induced by the field.  We treat the optical electric field in a unit 
cell as a time dependent, spatially-uniform electric field.

We consider the following time dependent Sch\"odinger equation for 
the Bloch orbitals to describe electron dynamics,
\be
i\hbar \frac{\partial}{\partial t}u_{n\vec{k}}(\vec{r},t)
=h\left[
\vec{k}+\frac{e}{\hbar c}\vec A(t)
\right]
u_{n\vec{k}}(\vec{r},t), 
\label{eq:TDSE} 
\ee
where $u_{n\vec{k}}(\vec{r},t)$ is a time dependent Bloch orbital
which is periodic in space, $u_{nk}(\vec r+\vec a)=u_{nk}(\vec r)$, 
where $\vec a$ is a lattice vector. The Hamiltonian 
$h[\vec{k}+e\vec{A}(t)/\hbar c]$ is assumed to have the following 
form: 
\be
h\left[
\vec{k}+\frac{e}{\hbar c}\vec A(t)
\right]=\frac{1}{2m}\left(
\vec{p}+\hbar \vec{k} +\frac{e}{c}\vec{A}(t)
\right)^2+V(\vec{r}) 
\label{eq:hamiltonian}
\ee
where $\hbar \vec{k}$ is a crystalline momentum.
$\vec{A}(t)$ is a time dependent, spatially-uniform vector potential.
It is related to the optical electric field $\vec E(t)$ by
$\vec{E}(t)=-\frac{1}{c}\frac{\partial}{\partial t}\vec{A}(t)$.
$V(\vec{r})$ is an effective single-electron potential which is periodic
in space.  It may be nonlocal in space to include, 
for example, a norm-conserving pseudopotential and/or an exchange potential.
When we describe the many-electron dynamics by the time-dependent 
Hartree-Fock or the TDDFT, we need to allow the potential $V(\vec r)$ 
to be time dependent.

\subsection{Choices of basis functions}

To calculate the electron dynamics in practice, we need to express 
Eq. (\ref{eq:TDSE}) in a matrix form.
In the real-space grid representation, we discretize the equation 
using uniform grids in the three-dimensional
(3D) Cartesian coordinates and express the Hamiltonian of 
Eq. (\ref{eq:hamiltonian}) as a sparse Hermitian matrix \cite{be00}. 
We call it the 3D grid representation hereafter.
Though this representation is simple and straightforward, 
it requires large computational resources to carry out simulations 
for a long period (typically a few tens of thousand time steps
for a few tens of femtosecond). 
Therefore, it is desirable to find representations which require
less computational resources without harming accuracy of the results. 
In this paper, we consider basis expansion methods converting the 3D
grid representation into dense matrix representations of smaller
dimensions. We will consider a few choices of basis functions and
examine their performances.

To convert the 3D grid representation into a dense matrix 
representation, a simple and natural choice may be to adopt a finite 
set of eigenstates of the static Hamiltonian
as a basis set to express the time-dependent orbital 
wave functions. For each $k$-point, we prepare occupied and 
unoccupied orbitals by solving,
\be
h[\vec k]v_{n\vec{k}}(\vec{r})
=\epsilon_{n\vec{k}}v_{n\vec{k}}(\vec{r}).
\label{eq:SKSeq}
\ee
We then expand the Bloch orbitals $u_{n \vec k}(\vec r,t)$ in terms of the
eigenfunctions $v_{m \vec k}({\vec r})$,
\be
u_{n\vec{k}}(\vec{r},t) 
= \sum_m c_{m}^{n\vec{k}}(t)v_{m\vec{k}}(\vec{r}),
\label{eq:k-fixed}
\ee
where $c_{m}^{n\vec{k}}(t)$ is a time-dependent coefficient 
for the expansion. We call it the $k$-fixed basis expansion.
If we includes all eigenfunctions in the expansion, 
it gives the same result as that in the 3D grid representation, 
since they constitute a complete set.
Truncating the unoccupied orbitals, we obtain an approximate matrix 
equation for the coefficients $\{ c_{m}^{n\vec{k}(t)}\}$.
However, as will be shown later, the convergence with respect to 
the number of unoccupied orbitals is very slow for some observables 
such as the current induced by the optical electric field, 
which in turn results in the slow convergence for the real part 
of the dielectric function.

As an alternative basis set, Houston states have been
extensively investigated and utilized to calculate
time evolutions of orbitals \cite{ho40,kr86}. The Houston states 
are defined as eigensfunctions of the single-electron Hamiltonian 
$h[\vec k+e\vec A(t)/\hbar c]$ 
at each time. They are given by
\be
v_{m \vec k}^H(\vec r,t)=v_{m \vec k+e \vec A(t)/\hbar c}(\vec r)
\exp \left[ -\frac{i}{\hbar} \int^t dt' \epsilon_{n \vec k + e\vec A(t')/\hbar c} \right],
\label{eq:Houston}
\ee
and are regarded as exact solutions in the adiabatic limit that is
the limit of slow variation of the electric field, $\vec E(t)$.
The Houston states have been used to investigate Bloch oscillations, 
Wannier-Stark ladders, and so on \cite{ho40,kr86,bl29,wa59}.
Using the Houston states, we may expand the Bloch orbitals as follows,
\be
u_{n\vec{k}}(\vec{r},t) 
= \sum_m c_{m}^{n\vec{k}}(t)v^H_{m\vec{k}}(\vec{r},t).
\ee
We may expect a fast convergence of the calculation for processes 
in which adiabatic approximation is well satisfied.

In practice, however, there are difficulties in preparing the Houston 
states in the first-principles descriptions. For example, consider
a case when two orbitals, one occupied and the other unoccupied,  
are (almost) degenerate at a certain $k$-point. When the
crystalline momentum with a shift of vector potential, 
$\vec k + e\vec A(t)/\hbar c$, goes through the crossing point,
it is difficult to decide which orbital should be occupied beyond
the point. Because of the difficulty, we will not explore the expansion
using the Houston states. Instead we will consider an alternative expansion
which effectively takes account of important aspects of the Houston states.

The essence of the expansion using the Houston states is a use of orbitals 
with crystalline momentum shifted by the external field.
This indicates that we should incorporate orbitals of shifted 
$k$-points in the basis functions, in addition to the orbitals of the 
original $k$-point. Of course, a set of all occupied 
and unoccupied orbitals for any $k$-points spans a complete set.
Therefore, any orbitals of a different $k$-point may eventually 
be expressible using eigenfunctions of the original $k$-point. 
However, inclusion of $k$-shifted orbitals remarkably accelerates 
the convergence of the expansion, as will be shown later.

Our prescription to use the orbitals with different $k$-points is
summarized as follows. To express time-dependent Bloch orbitals 
$u_{n\vec k}(\vec r,t)$, we employ occupied orbitals of the 
Hamiltonian in which crystalline momenta are shifted from the 
original $k$-point, $\vec k+e\Delta \vec A/\hbar c$, in 
addition to occupied and unoccupied orbitals of the original $k$-point. 
We call it the $k$-shifted basis expansion.
We expect the Houston states $v^H_{n \vec k}(\vec r,t)$ may be
described well by a superposition of orbitals of the original 
$k$-point, $v_{n \vec k}(\vec r)$, and those of shifted $k$-points, 
$v_{n \vec k+e\Delta \vec A/\hbar c} (\vec r)$. 
We note that orbitals of different $k$-points are no more 
orthogonal to each other. In our practical calculations, we first 
prepare an orthonormalized basis set from orbitals of different
$k$-points.

In the following section, we will examine performance of 
the $k$-shifted basis expansion for electron dynamics calculations.
For comparison, we will compare results of $k$-shifted basis
expansion with those of $k$-fixed basis expansion, and
with those without any basis expansions, namely, direct
calculations in the 3D grid representation.

\section{Linear response}

To test the basis expansion methods, we calculate
a dielectric function of $\alpha$-quartz using a density functional Hamiltonian.
We calculate the dielectric function from a real time
evolution of orbitals employing $k$-fixed and $k$-shifted basis 
expansions as well as 3D grid representation without any
basis expansions. We note that responses to any weak external fields 
are fully characterized by the dielectric function.

We use a first-principles density functional Hamiltonian in the 
local density approximation (LDA) \cite{pe81}. 
We denote the static Kohn-Sham Hamiltonian as 
\begin{equation}
h_{KS}=\frac{p^2}{2m}+V_{loc}+V_{nonloc},
\end{equation}
where $V_{loc}$ is a local potential composed of electron-ion and
electron-electron interactions. $V_{nonloc}$ is a nonlocal electron-ion
interaction in the pseudopotential description \cite{tr91} with
a separable approximation \cite{kl82}.
We ignore local field effects which may be included in the TDDFT. 
We have confirmed that the local field effects are very small in 
the TDDFT under the adiabatic LDA.
It is well-known that the dielectric functions 
in the LDA are not accurate enough to reproduce measured
features. However, since our purpose here is to compare efficiencies 
of different numerical methods, we consider it is sufficient.

The time evolution of Bloch orbitals is described by
\begin{eqnarray}
&i&\hbar \frac{\partial}{\partial t} u_{nk}(r,t)=
\left [
\frac{1}{2m} \left(
\vec{p}+\hbar \vec{k}+\frac{e}{c}\vec{A}(t)
\right)^2 +V_{loc} 
+e^{-i(\hbar \vec{k} +e\vec{A}(t)/c)\cdot \vec{r}} \hat{V}_{nonloc} 
e^{i(\hbar \vec{k} +e\vec{A}(t)/c)\cdot \vec{r}} 
\right ] u_{nk}(r,t),
\nonumber \\
\label{eq:TDSE2}
\end{eqnarray}
where $\vec A(t)$ is a vector potential which is related
to the applied electric field $\vec E(t)$ by
$\vec E(t)=-\frac{1}{c} \frac{d\vec A(t)}{dt}$.
Optical properties can be extracted from the current
averaged over the unit cell, which is induced by the applied electric 
field,
\be
\vec{J}(t)=-\frac{e}{\Omega}\sum_{n,\vec{k}} 
\int_{\Omega} d\vec{r} d\vec{r'} \ u^*_{n\vec{k}}(\vec{r},t)  
e^{-i( \vec{k} +e\vec{A}(t)/\hbar c)\cdot \vec{r}} 
\vec{v}(\vec{r},\vec{r'})
e^{i( \vec{k} +e\vec{A}(t)/\hbar c)\cdot \vec{r'}} 
u_{n\vec{k}}(\vec{r'},t),
\label{eq:macro_current}
\ee
where $\Omega$ is the volume of the unit cell and the velocity operator 
$\vec{v}(\vec{r},\vec{r'})$ is defined by
\be
\vec{v}(\vec{r},\vec{r'})
=
\frac{-i\hbar}{m} \vec{\nabla} \delta(\vec{r}-\vec{r'})
+\frac{1}{i\hbar} \left[
\vec{r}\hat{V}_{nonloc}(\vec{r},\vec{r'})
-\hat{V}_{nonloc}(\vec{r},\vec{r'})\vec{r'}
\right].
\label{eq:kin-momentum}
\ee
For a weak field, the applied electric field and the induced current 
are linearly related by the conductivity. 
Assuming the field has only a $\beta$-direction component, 
the frequency-dependent conductivity $\sigma_{\alpha \beta}(\omega)$ 
and the dielectric function $\epsilon_{\alpha \beta}(\omega)$ are 
given as follows:
\be
\sigma_{\alpha \beta} (\omega) = \frac{\int_{-\infty}^{\infty} dt 
e^{i\omega t} J_{\alpha}(t)} 
{\int_{-\infty}^{\infty} dt e^{i\omega t} E_{\beta}(t)},
\hspace{5mm}
\epsilon_{\alpha\beta} (\omega) = \delta_{\alpha\beta} 
+\frac{4\pi i \sigma_{\alpha\beta}(\omega)}{\omega}.
\label{eq:sgm_eps}
\ee

For any choices of the electric field, $E_{\beta}(t)$,
we may obtain the same results for the conductivity and the 
dielectric function so long as the field used in the calculation 
is sufficiently weak.
For computational convenience, we employ
an impulsive electric field, $E_{\beta}(t) = d\delta(t)$,
where $d$ is a parameter specifying the intensity of the field.
This choice is equivalent to a constant shift of the vector 
potential, $A_{\beta}(t) = -cd\theta(t)$. Using the impulsive 
field, the conductivity as a function of time, 
$\sigma_{\alpha \beta}(t)$, is proportional to the induced current,
$J_{\alpha}(t)$. To reduce numerical noises which originate from 
a finite period of the calculation, we introduce a mask function 
in the Fourier transformation,
\begin{equation}
\sigma_{\alpha\beta}(\omega)= \frac{1}{d} 
\int_0^T dt e^{i\omega t} J_{\alpha}(t) W(t/T),
\end{equation}
where $T$ is the period of the time evolution. For the mask
function $W(x)$, we employ $W(x)=1-3x^2+2x^3$.

For the calculations of SiO$_2$, we choose a unit cell of rectangular 
parallelepiped containing six silicon and twelve oxygen atoms,
96 valence electrons in total. We note that the number of occupied 
orbitals at each $k$-point is 48 since all orbitals are doubly 
occupied. We first calculate the ground state
using a uniform grid representation in the 
3D Cartesian coordinates. 
We choose $z$-axis parallel to the $c$-axis of the $\alpha$-quarts,
and discretize the $x$, $y$, and $z$ coordinates of the unit cell 
into 20, 36 and 50 grid points, respectively.
For the Brillouin zone, we sample it using $4 \times 4 \times 4$ grid 
points. Using the 3D grid representation, we can directly calculate 
the time propagation of the orbitals using, for example, 
the Taylor expansion method \cite{ya96,fl78}. 
We will call it the 3D grid calculation. We also use 
the 3D grid representation to calculate occupied and unoccupied 
orbitals in the ground state by solving Eq. (\ref{eq:SKSeq}). 
These orbitals will be used in the $k$-fixed and $k$-shifted basis 
expansions. We use the Taylor expansion method for 
all time-propagation calculations of $k$-fixed, $k$-shifted basis
expansions as well as 3D grid calculations.

\subsection{$k$-fixed basis expansion}

\begin{figure}[tb]
 \begin{center}
  \includegraphics[width= 8cm]{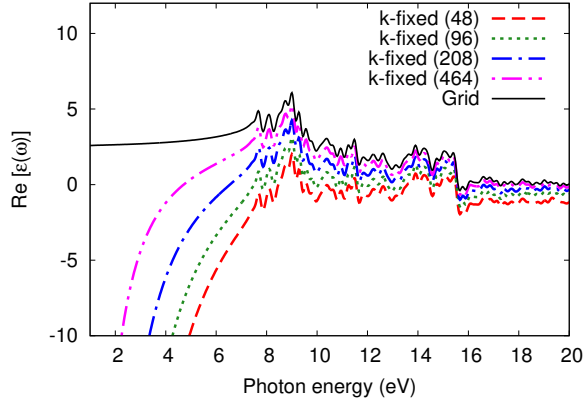}
 \end{center}

 \caption{Real part of the dielectric function of $\alpha$-quartz 
calculated by using several $k$-fixed basis expansions. 
The number in the parenthesis of the legends represents 
the number of unoccupied orbitals employed at each $k$-point. 
Black-solid line shows results of the 3D grid calculation.}
 \label{fig:eps_Re_kfixed}
\end{figure}

We first examine the $k$-fixed basis expansion described by
Eq. (\ref{eq:k-fixed}).
Figure \ref{fig:eps_Re_kfixed} shows the real part of the 
dielectric function of $\alpha$-quartz using $k$-fixed basis 
expansions with various truncations of unoccupied orbitals 
as well as that in the 3D grid calculation. In the figure,
$k$-fixed basis expansions are denoted as $k$-fixed ($N_{unocc}$), 
where $N_{unocc}$ is the number of unoccupied orbitals employed in 
the expansion at each $k$-point.
For example, $k$-fixed ($464$) means employment of $48$
occupied and $464$ unoccupied orbitals.
We note that, in the calculation of $48$ occupied and $464$ unoccupied
orbitals, the matrix dimension of the Hamiltonian is 512. This is much 
smaller than the dimension in the 3D grid representation which is equal
to the number of the grid points, $20 \times 36 \times 50 = 36000$.

In the figure, one can find that the real part of the dielectric 
function using the $k$-fixed basis expansion shows negative 
divergent behavior which is absent in the 3D grid calculation.
The divergent behavior diminishes as the number of 
unoccupied orbitals employed in the expansion increases. 
However, the convergence is rather slow. Even when the $k$-fixed 
basis expansion includes $512$ orbitals in total, 
the dielectric function in the $k$-fixed basis expansion is much 
different from that of the 3D grid calculation.

\begin{figure}[htbp]
 \begin{center}
  \includegraphics[width= 8cm]{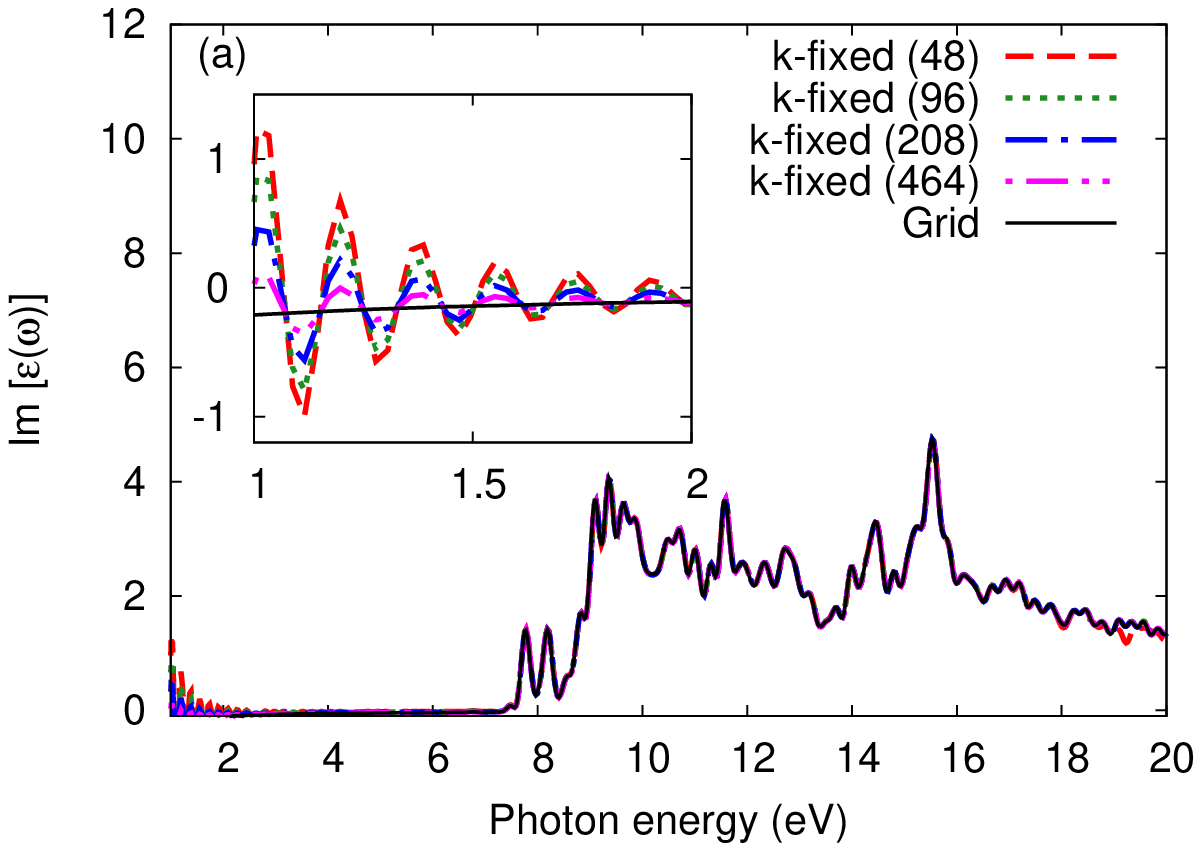}
  \includegraphics[width= 8cm]{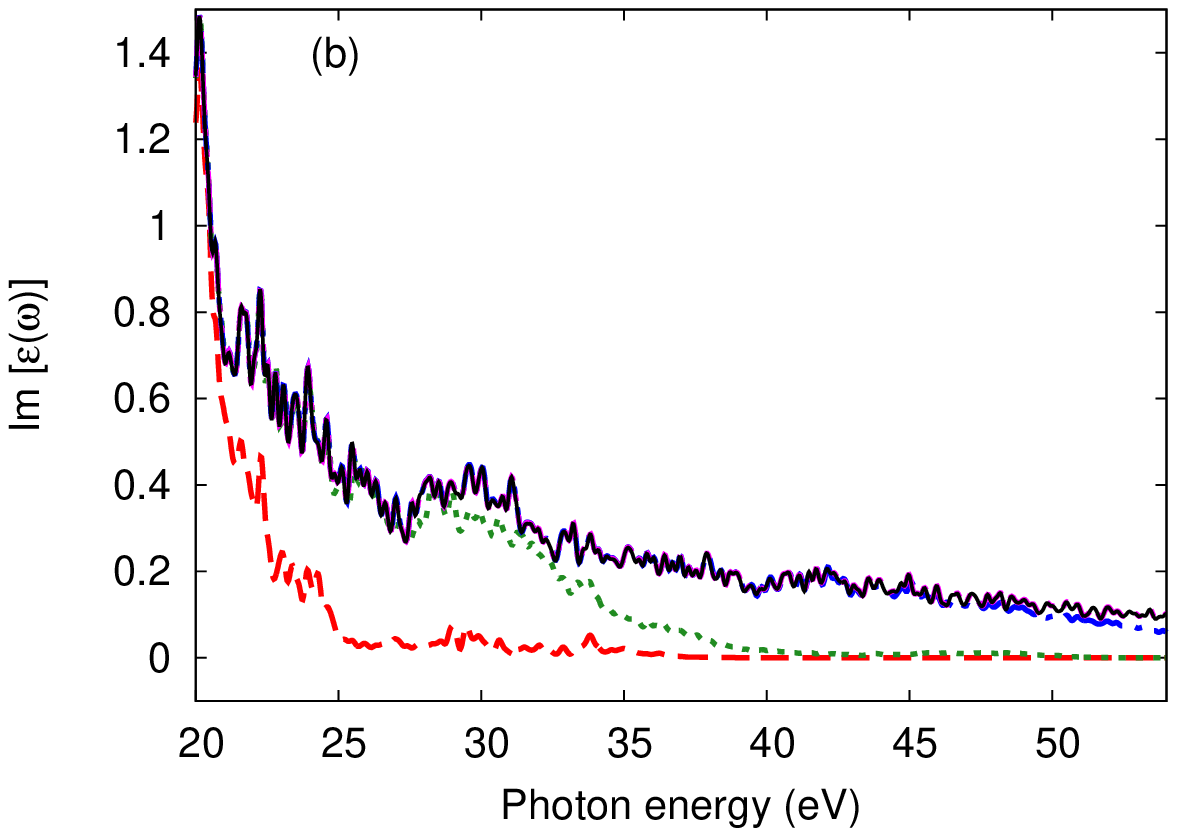}
 \end{center}

 \caption{Imaginary part of dielectric functions of 
$\alpha$-quartz calculated by using several $k$-fixed basis expansion.
The number in the parenthesis of the legends represents 
the number of unoccupied orbitals employed at each $k$-point. 
Black-solid line shows results of the 3D grid calculation.
The panel (a) shows lower energy region from 1 eV to 20 eV,
and the panel (b) shows higher energy region from 20 eV to 54 eV.}
 \label{fig:eps_Im_low_kfixed}
\end{figure}

Figure \ref{fig:eps_Im_low_kfixed} (a) shows the imaginary part of 
the dielectric function below 20 eV, while 
Figure \ref{fig:eps_Im_low_kfixed} (b) shows that above 20 eV.
The imaginary part below 20 eV is well described by
the $k$-fixed basis expansion with any number of unoccupied orbitals, 
except below 2 eV. The $k$-fixed basis 
expansion calculation brings unphysical oscillations below 2eV, 
which are absent in the 3D grid calculation. 
This oscillation is sensitive to the number of unoccupied 
orbitals used in the $k$-fixed basis expansion and 
diminishes with increasing the number of unoccupied 
orbitals. We next look at responses at higher energy.
As seen from the panel (b), the result of the 
$k$-fixed basis expansion becomes poor as 
the excitation energy increases. This can be explained by the truncation 
of the basis functions.
The energy difference between the highest 
occupied orbital and the highest unoccupied orbital in the truncations is 
given approximately by 19 eV for the $k$-fixed (48), 
31 eV for the $k$-fixed (96), 51 eV for the $k$-fixed (208),
and 92 eV for the $k$-fixed (464) basis expansions.
In Fig. \ref{fig:eps_Im_low_kfixed} (b), one can find 
a deviation between the 3D grid calculation and the $k$-fixed basis 
expansions starts to occur at the corresponding energies. 
For example, results of the $k$-fixed (48) basis expansion and of
the 3D grid calculation differ much above 21 eV, which roughly coincides 
with the energy difference of 19 eV between the highest occupied 
orbitals and the highest unoccupied orbitals included in the $k$-fixed (48) 
basis expansion.

\subsection{k-shifted basis expansion}

We next examine the $k$-shifted basis expansion, adding occupied 
orbitals of nearby $k$-points to the basis.
Since the magnitude of the vector potential, $\vec A(t)$, may be 
infinitesimally small in linear response calculations, we may use 
a small value for the $k$-shifts. In practice, for responses in 
$z$-direction, we employ occupied orbitals at shifted $k$-points of
$e \Delta\vec{A}/\hbar c=(0,0,0.01)$ and at 
$e \Delta \vec{A}/\hbar c=(0,0,-0.01)$ in atomic unit. 
We denote the $k$-shifted basis expansion as $k$-shifted-2 
($N_{unocc}$), where the number 2 indicates that we employ 
occupied orbitals at two shifted $k$-points, 
$(0,0,\pm 0.01)$ in the present case, and $N_{unocc}$ represents 
the number of unoccupied orbitals at the original $k$-point.
For example the $k$-shifted-2 ($144$) basis expansion includes
$144$ unoccupied orbitals at the original $k$-point as well as
$48\times 3$ occupied orbitals at the original $k$-point and two 
shifted $k$-points. Thus, it includes 288 orbitals in total.

\begin{figure}[tb]
 \begin{center}
  \includegraphics[width= 8cm]{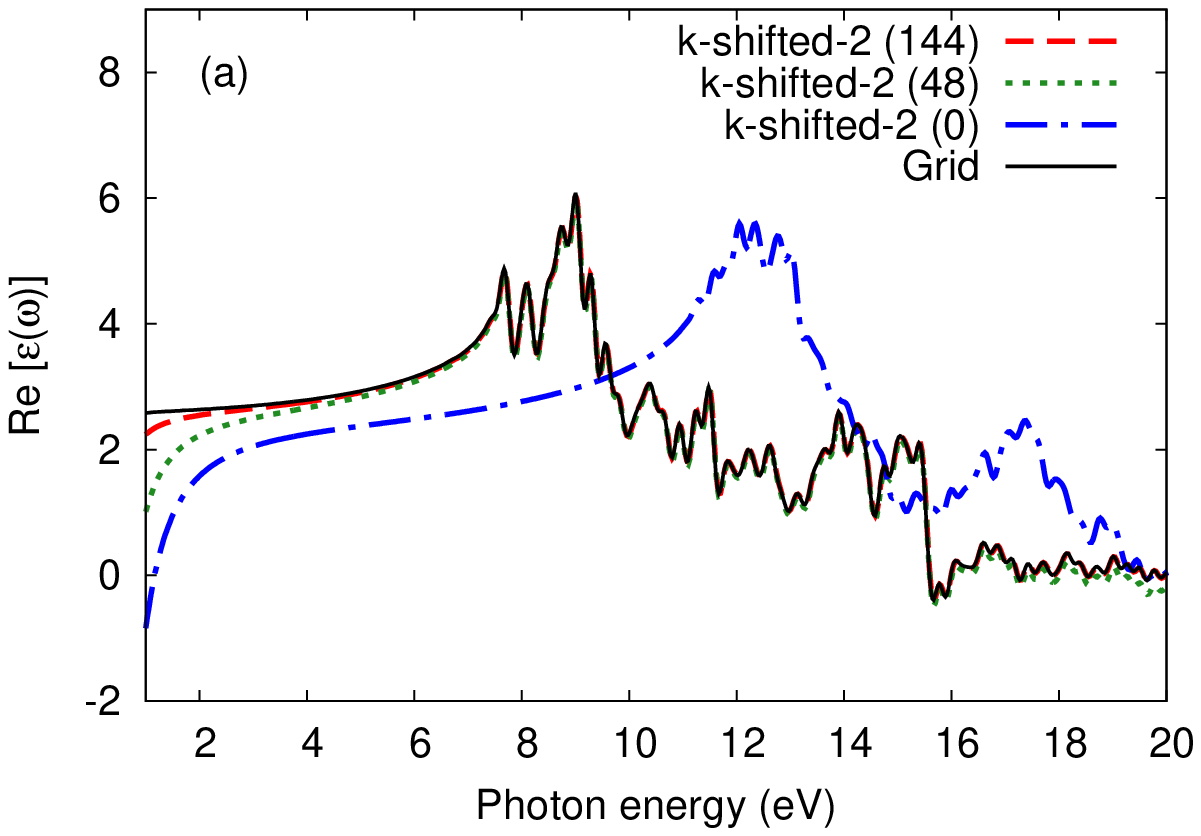}
  \includegraphics[width= 8cm]{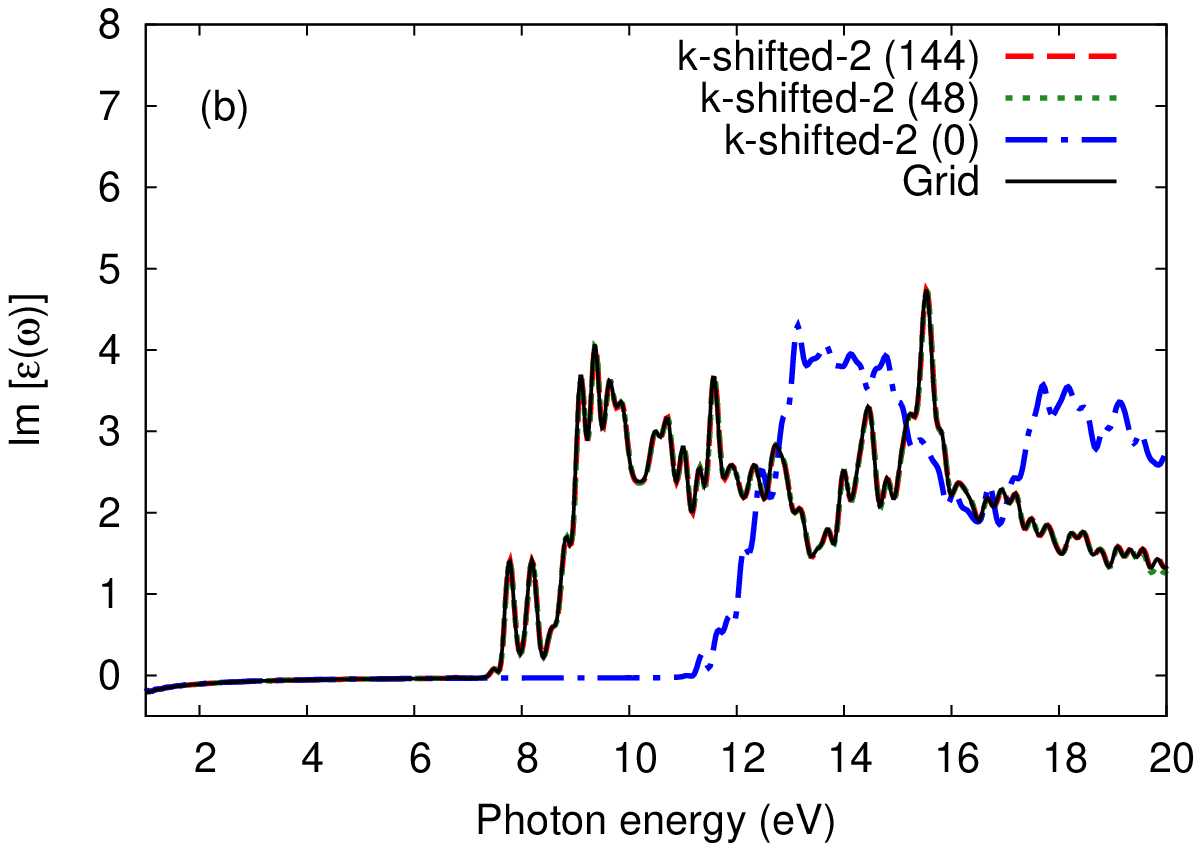}
 \end{center}

 \caption{Real (a) and imaginary (b) parts of the dielectric function 
of $\alpha$-quartz calculated by using $k$-shifted basis expansions.
The number in the parentheses of legends represents the number of 
unoccupied orbitals employed at original $k$-point. The result of 
the 3D grid calculation is also shown.}
 \label{fig:eps_Re_kshifted}
\end{figure}

Figure \ref{fig:eps_Re_kshifted} shows dielectric function 
using the $k$-shifted basis expansions 
of different numbers of unoccupied orbitals, and that of the
3D grid calculation. As seen from the figure, the $k$-shifted-2 (48) 
and (144) basis expansions well reproduce both real and 
imaginary parts of the dielectric function in the 3D grid calculation. 
Since the real part of the dielectric function cannot be described 
in the $k$-fixed basis expansion as shown 
in Fig. \ref{fig:eps_Re_kfixed}, the present result clearly
indicates that the inclusion of the $k$-shifted orbitals
is essential to describe the linear response accurately.

In Fig. \ref{fig:eps_Re_kshifted}, we find the $k$-shifted basis 
expansion without any unoccupied orbitals of original $k$-point, 
the blue-dash-dotted curve of $k$-shifted-2 
(0), fails to reproduce either real or imaginary parts of 
the dielectric function. This fact indicates that 
inclusion of both occupied orbitals of shifted $k$-point
and unoccupied orbitals of original $k$-point are indispensable
to correctly describe the liner response.

Although we here calculate the dielectric function using the real time
method, the importance of including $k$-shifted occupied orbitals
in linear response calculations should also be true in other 
computational methods such as the Sternheimer
method in which linear response for a given frequency is calculated 
solving linear algebraic equations.

\subsection{Intra-band and inter-band transitions}

In this subsection, we present an analytical 
investigation on the reason why the inclusion of $k$-shifted occupied orbitals 
brings fast convergence for the calculation of the dielectric function.

We start with the following formal solution of Eq. (\ref{eq:TDSE2}):
\be
u_{n\vec k}(\vec r,t)=\hat{T} \exp \left[-\frac{i}{\hbar} 
\int^t_{t_0}dt' h[\vec k+e\vec A(t')/\hbar c]  \right]
u^{GS}_{n\vec k}(\vec r),
\ee
where $\hat T$ is the time ordering operator.
As we described below Eq. (\ref{eq:sgm_eps}), we consider an impulsive 
distortion given by $\vec E=-\vec s \hbar \delta(t)/e$ with 
$\vec s =s \vec e_{\beta}$, 
where $s$ is an infinitesimally small parameter and 
$\vec e_{\beta}$ is a unit vector in $\beta$ direction.
For this distorting field, the Hamiltonian 
$h[\vec k+e\vec A(t)/\hbar c] $ can be described as follows:
\be
h[\vec k+e\vec A(t)/\hbar c] =\frac{1}{2m}\left (
\vec p + \hbar \vec k + \hbar \vec s \Theta (t) 
\right )^2+V =
\left\{ \begin{array}{ll}
    h[\vec k+\vec s] & (t>0) \\
    h[\vec k] & (t \le 0). 
  \end{array} \right.
\ee

The induced current averaged over the unit cell is given by 
Eq. (\ref{eq:macro_current}). 
We decompose the current into orbitals,
$\vec J(t) = \sum_{n\vec k} \vec J_{n\vec k}(t)$.
We define a change of the current of each orbital by
\begin{eqnarray}
&&  \delta \vec J_{n\vec k}(t) =
\left(  \vec J_{n\vec k}(t)- \vec J_{n\vec k}(t=0) \right)  \nonumber \\
&=& -\frac{e}{\Omega} \left[ 
\langle u^{GS}_{n\vec k} | e^{i h[\vec k+\vec s] t /\hbar  }
e^{-i(\vec{k}+\vec{s})\cdot \vec r} \vec{v} e^{i(\vec{k}+\vec{s})\cdot \vec r}
e^{-i h[\vec k+\vec s] t /\hbar } | u^{GS}_{n\vec k}\rangle -\langle u^{GS}_{n\vec k} 
| e^{-i\vec{k} \cdot \vec r} \vec{v} e^{i\vec{k} \cdot \vec r} | u^{GS}_{n\vec k}\rangle \right].
\label{eq:curr01}
\end{eqnarray}

Applying the first order perturbation theory to Eq. (\ref{eq:curr01})
assuming a small distortion amplitude of $s$, the $\alpha$-component 
of the current can be expressed as follows:
\begin{eqnarray}
\delta J^{\alpha}_{n\vec k}(t) &=&
-s \frac{e}{\Omega\hbar}
\frac{\partial^2 \epsilon_{n\vec k}}{\partial k_{\alpha} \partial k_{\beta}}
+ s \frac{e\hbar}{\Omega}\sum_{n' \neq n} \frac{
\langle u^{GS}_{n\vec k} | e^{-i\vec{k}\cdot \vec r}
v^{\alpha} e^{i \vec{k}\cdot \vec r)} | u^{GS}_{n'\vec k} \rangle 
\langle u^{GS}_{n'\vec k} | e^{-i \vec{k}\cdot \vec r}
v^{\beta} e^{i \vec{k} \cdot \vec r} | u^{GS}_{n\vec k} \rangle 
}
{\epsilon_{n\vec k}-\epsilon_{n' \vec k}} \nonumber \\
&\times&\left \{
e^{i(\epsilon_{n \vec k}-\epsilon_{n' \vec k})t/\hbar}
+e^{-i(\epsilon_{n \vec k}-\epsilon_{n' \vec k})t/\hbar}
\right \},
\label{eq:curr02} 
\end{eqnarray}
where $\delta J^{\alpha}_{n\vec k}$ and $v^{\alpha}$ 
represent $\alpha$-components of $\delta \vec J_{n \vec k}$
and $\vec v$, respectively. The first term of the right 
hand side is defined as follows:
\be
\frac{\partial ^2\epsilon_{n\vec k}}{\partial k_{\alpha}\partial k_{\beta}}
= \hbar
\lim_{s \rightarrow 0}
\frac{1}{s}\left[ 
\langle u^{GS}_{n\vec k+\vec s} | e^{-i(\vec{k}+\vec{s}) \cdot \vec r}v^{\alpha}e^{i(\vec{k}+\vec{s}) \cdot \vec r}| 
u^{GS}_{n\vec k+\vec s} \rangle 
-\langle u^{GS}_{n\vec k} | e^{-i\vec{k}\cdot \vec r}  v^{\alpha} e^{i\vec{k}\cdot \vec r} |u^{GS}_{n\vec k} \rangle 
\right].
\ee

The first term of Eq. (\ref{eq:curr02}) is a time-independent current 
induced by intra-band transitions. The second term is a time-dependent 
oscillating current induced by inter-band transitions. 
We thus call the first and the second terms intra-band and 
inter-band currents, respectively.

Applying the Fourier analysis of Eq. (\ref{eq:sgm_eps}) to 
the current of Eq. (\ref{eq:curr01}), one can obtain an analytic form 
of the dielectric function as follows:
\be
\epsilon_{\alpha \beta}(\omega)=
1+\frac{4\pi i}{\omega}\sum_{n \vec k} \sigma^{n \vec k}_{\alpha \beta},
\ee
where $\sigma^{n \vec k}_{\alpha \beta}$ is a conductivity 
decomposed into orbitals and $k$-points. They are given by 
\begin{eqnarray}
\sigma^{n \vec k}_{\alpha \beta} (\omega) 
&=&
i\frac{e^2}{\Omega \hbar ^2}\frac{\partial^2 \epsilon_{n \vec k}}
{\partial k_{\alpha} \partial k_{\beta}}
\frac{1}{\omega+i\gamma} 
-i\frac{e^2}{\Omega}\sum_{n' \neq n} \frac{
\langle u^{GS}_{n \vec k} | e^{-i\vec{k}\cdot \vec r} v^{\alpha} e^{i\vec{k}\cdot \vec r}| u^{GS}_{n' \vec k} \rangle 
\langle u^{GS}_{n' \vec k} | e^{-i\vec{k}\cdot \vec r} v^{\beta} e^{i\vec{k}\cdot \vec r}| u^{GS}_{n \vec k} \rangle 
}
{\epsilon_{n' \vec k}-\epsilon_{n \vec k}} 
\nonumber \\
&\times& 
\left[
\frac{1}{\epsilon_{n' \vec k}-\epsilon_{n \vec k}-\omega-i\gamma}
+\frac{1}{\epsilon_{n \vec k}-\epsilon_{n' \vec k}-\omega-i\gamma}
\right]. \nonumber \\
\label{eq:sigma-an}
\end{eqnarray}
Here we added a small imaginary quantity $\gamma$ in the frequency.
We note that the second derivative of the single particle 
energy is proportional to the inverse of the effective mass of electrons 
$\partial ^2 \epsilon_{n \vec k}/\partial k_{\alpha}\partial_{\beta} 
= \hbar ^2 /m^*_{\alpha \beta}$.
In Eq. (\ref{eq:sigma-an}), the intra-band transitions contribute to 
the first term which shows a Drude behavior, and the inter-band 
transitions contributes to the second term.

For insulators, the first Drude term of Eq. (\ref{eq:sigma-an}) should be 
identically zero after integrating it over $k$.
In the $k$-fixed basis expansion, however, we found a Drude-like
contribution in the real part of the dielectric function 
as seen in Fig.\ref{fig:eps_Re_kfixed}.
This indicates that the slow convergence in the $k$-fixed basis expansion 
comes from a failure to cancel the intra-band transitions which 
should vanish if one includes all unoccupied orbitals.

\begin{figure}[tb]
 \begin{center}
  \includegraphics[width= 8cm]{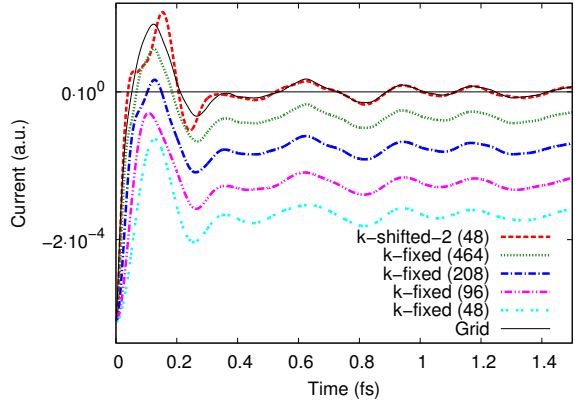}
 \end{center}

 \caption{Currents as functions of time induced by an instantaneous distortion at $t=0$. 
Results of $k$-fixed basis expansions, $k$-shifted basis expansion, and the 3D grid calculation are shown. }
 \label{fig:current_eps}
\end{figure}

Figure \ref{fig:current_eps} shows electric currents as functions of 
time after the distortion is applied at $t=0$.
The currents using the $k$-fixed basis expansion, the $k$-shifted basis 
expansion, and the 3D grid calculation are shown. 
These currents are used in the calculations of dielectric functions shown
in Fig \ref{fig:eps_Re_kfixed}, \ref{fig:eps_Im_low_kfixed}, and 
\ref{fig:eps_Re_kshifted}.

As expected, one sees that large constant components appear in 
the currents of the $k$-fixed basis expansions.
As the number of unoccupied orbitals increases in the $k$-fixed basis 
expansion, the constant component of the current decreases. 
However, the constant component remains substantial even when 
a large number of unoccupied orbitals, 464 orbitals, is employed in 
the $k$-fixed (464) basis expansion. In contrast, the current
of the $k$-shifted basis expansion oscillates around zero-value
without any constant components and coincides well with 
the result of the 3D grid calculation.

Summarizing the findings in this subsection,
the $k$-fixed basis expansion fails to describe the current caused
by the intra-band transitions which should vanish in any insulators.
Since the appearance of the constant current in the $k$-fixed basis expansion 
indicates that the fundamental property of insulators cannot be
described adequately, the $k$-fixed basis expansion should not be 
used for the electron dynamics calculations.
We have found the addition of the occupied orbitals of shifted $k$-points
to the basis functions of original $k$-point solves the problem.
The $k$-shifted basis expansion efficiently describes
the electron dynamics in insulators in the linear response regime.

\section{Electron dynamics induced by pulsed electric fields}

In this section, we examine usefulness of the $k$-shifted
basis expansion to describe electron dynamics under pulsed electric 
fields. In the following calculations, we employ the pulsed electric 
field described by the vector potential,
\be
\vec{A}(t)=\vec{e}_z \frac{cE_0}{\omega} \cos(\omega t) 
\sin^2\left( \frac{\pi t}{T_p} \right) \hspace{10mm} (0<t<T_p),
\label{eq:laser}
\ee
where $\omega$ represents the mean frequency of 
the pulse field for which we take $\omega=3.10$ eV$/\hbar$, and
$T_p$ represents a pulse duration for which we take 
$T_p=10.67$ fs corresponding to eight cycles.
We define the peak laser intensity of the pulsed electric field
by $I=c E_0^2/8\pi$.
We will consider electron dynamics in $\alpha$-quartz as before,
for two cases of different field intensities.

We calculate two physical quantities to examine quality of
the representations. One is the electric current, $\vec J(t)$, defined 
by Eq. (\ref{eq:macro_current}). The other is the electronic excitation energy 
of the system per unit cell,  $E_{ex}(t)$, defined as follows:
\be
E_{ex}(t)=\sum_{n \vec k} \int_{\Omega} d\vec r 
u^*_{n\vec{k}}(\vec{r},t) h \left[ {\vec k + \frac{e}{\hbar c} \vec A(t)} \right] u_{n\vec{k}}(\vec{r},t)
-E_{GS},
\ee
where $E_{GS}$ is the energy of the ground state.

\subsection{Electron dynamics under a weak pulsed electric field}

First we examine basis expansion methods applied for electron 
dynamics under a weak pulsed electric field.
The peak laser intensity is set to $I=1.0\times 10^{10}$ W/cm$^2$.
At this intensity, electron dynamics is well described by the linear
response theory. Since the frequency $\omega=3.10$ eV/$\hbar$ is 
below the optical band gap of $\alpha$-quartz which is 6.3 eV 
in the present LDA calculation, real electronic excitations do
not take place.

\begin{figure}[tb]
 \begin{center}
  \includegraphics[width= 8cm]{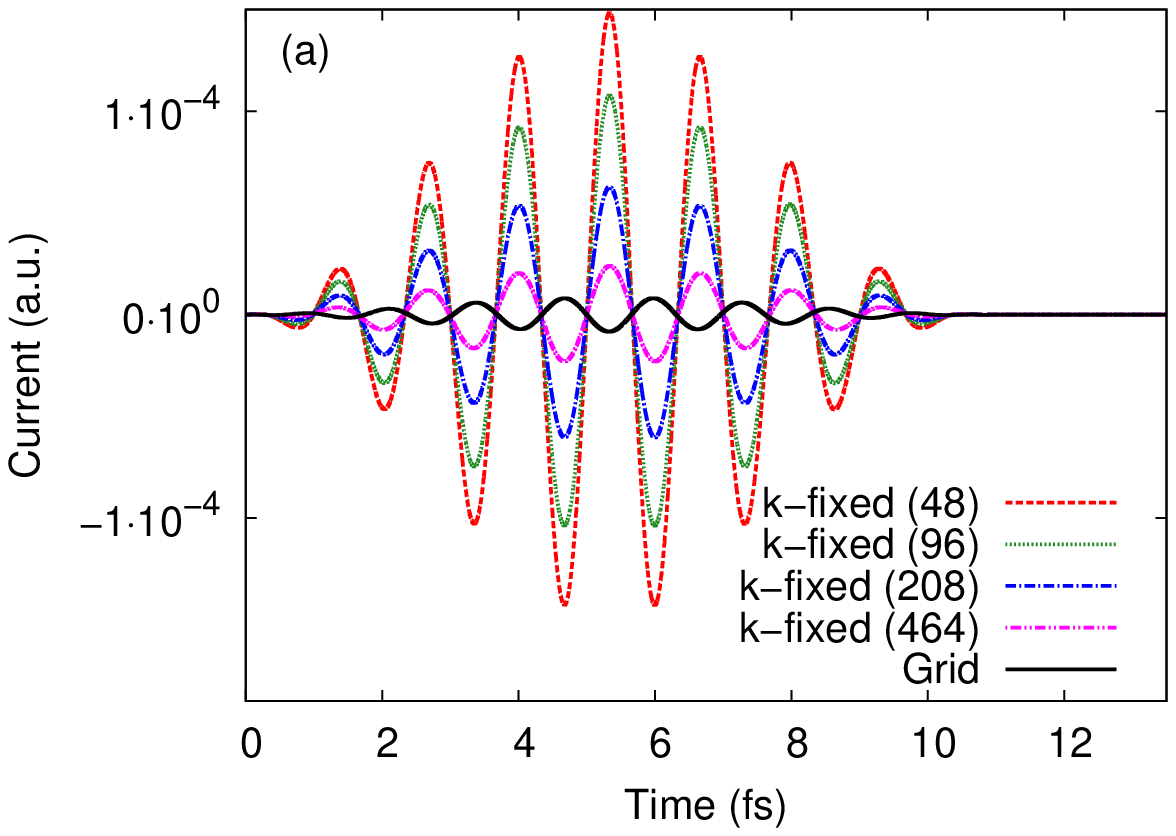}
  \includegraphics[width= 8cm]{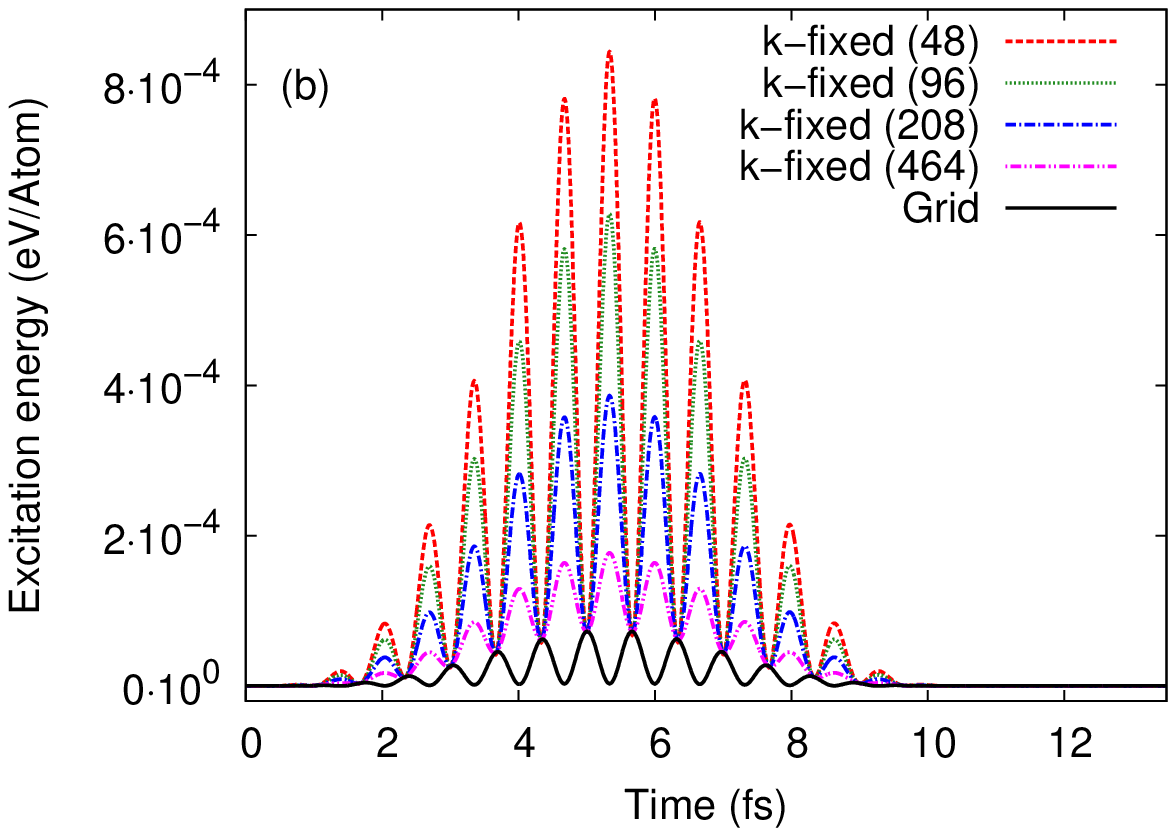}
 \end{center}

\caption{Currents (a) and excitation energy per unit cell (b) 
as functions of time are shown. 
Black-solid line shows results of the 3D grid calculation,
while other lines show results of the $k$-fixed basis expansions
with different truncations. The maximum intensity 
of the pulsed electric field is set to $I=1.0\times 10^{10}$ W/cm$^2$ }
 \label{fig:current_kfixed_p1d10}
\end{figure}

We show results using $k$-fixed basis expansions in 
Fig. \ref{fig:current_kfixed_p1d10}.
Figure \ref{fig:current_kfixed_p1d10} (a) shows
currents and (b) shows excitation energies per unit cell
as functions of time.
The black-solid lines represent results of the 3D grid
calculation, while other lines show results of $k$-fixed basis 
expansions. The number of unoccupied orbitals employed is denoted
in the parentheses.

As seen from the panel (a), calculations using $k$-fixed basis 
expansions cannot reproduce the current of the 3D grid calculation. 
Even the sign of the current is opposite between the $k$-fixed 
basis expansions and the 3D grid calculation.
The magnitude of the current in the $k$-fixed basis expansion 
is much larger than that in the 3D grid calculation,
even using the largest number of unoccupied orbitals, $k$-fixed (464). 
These observations are consistent with the dielectric function calculated 
using the $k$-fixed basis expansions. Namely, in Fig. \ref{fig:eps_Re_kfixed},
the real part of the dielectric function is 
negative at the frequency of, $\hbar \omega =3.1$ eV, in the $k$-fixed 
basis expansions, which is opposite in sign to the 3D grid 
calculation. The magnitude of the dielectric function decreases as 
the number of unoccupied basis functions increases.

In the panel (b), one can find that the 
excitation energy during the pulse field irradiation is always 
overestimated in the $k$-fixed basis expansions. 
The times when excitation energy shows maximum are
different between the 3D grid calculation and the $k$-fixed basis 
expansion. The energy maximum of the 3D grid calculation 
takes place when the magnitude of the electric field is maximum, while
the energy maximum in the $k$-fixed basis expansions takes place when
the magnitude of the vector potential shows maximum.
These facts indicate that calculations in the $k$-fixed basis 
expansions fail to reproduce the 3D grid calculation even in a 
qualitative level.

\begin{figure}[htbp]
 \begin{center}
  \includegraphics[width= 8cm]{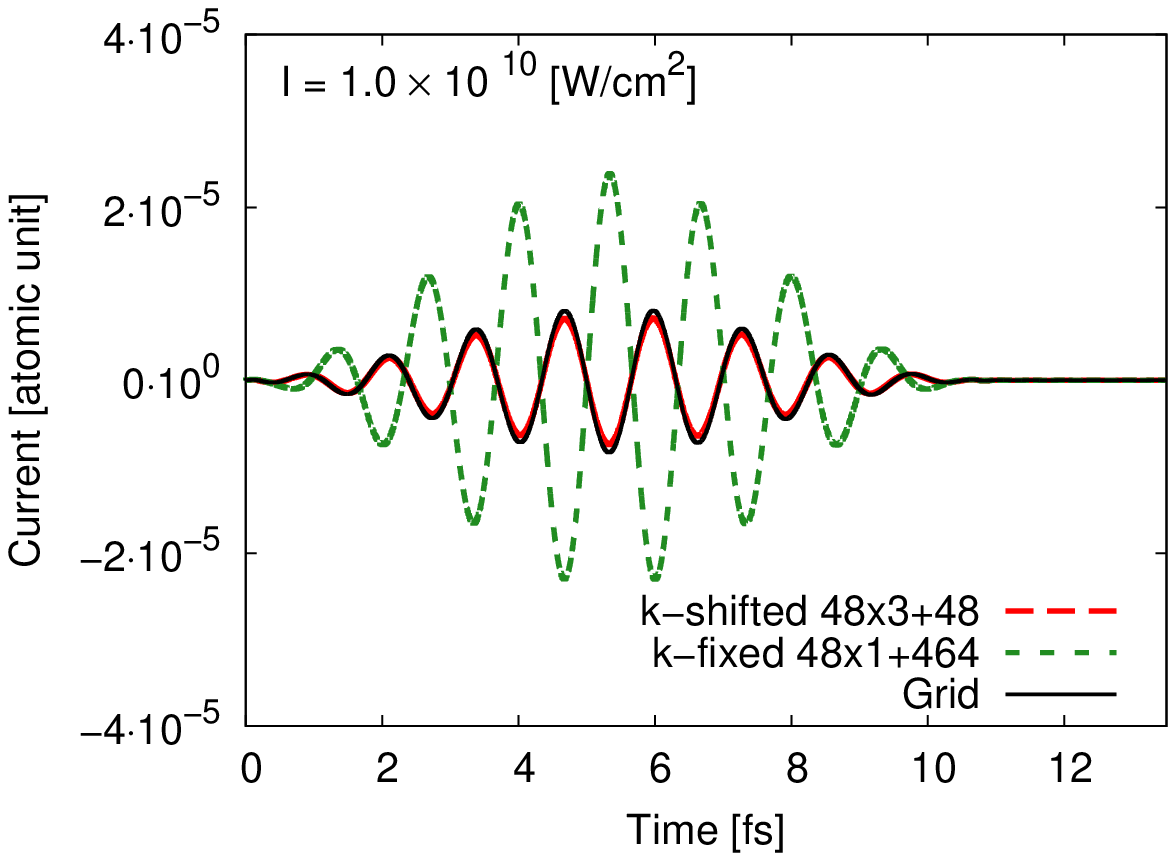}
  \includegraphics[width= 8cm]{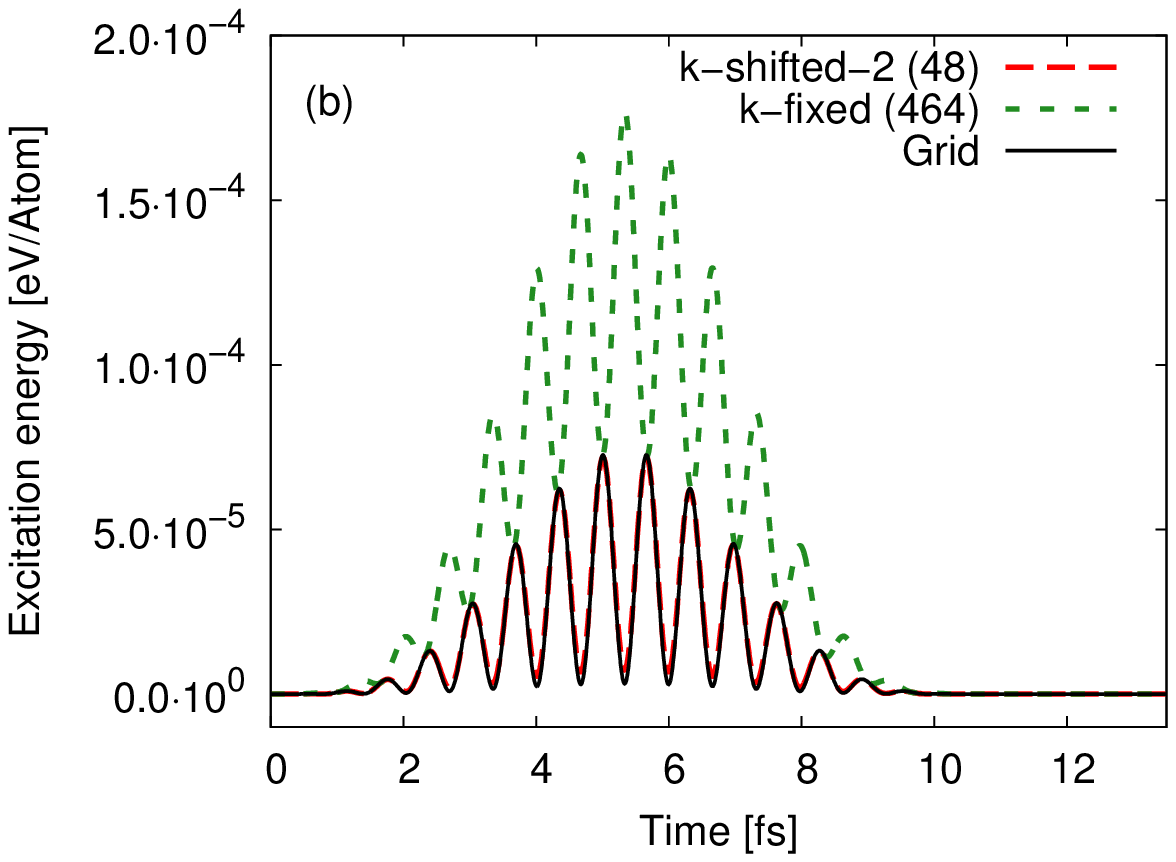}
 \end{center}

 \caption{Currents (a) and excitation energies per unit cell (b)
as functions of time are shown.
Black-solid line shows results of the 3D grid calculation,
red-dashed line shows results of the $k$-shifted basis expansion,
and green-dotted line shows results of the $k$-fixed basis expansion.
The maximum intensity of the pulsed electric field is set to 
$I=1.0\times 10^{10}$ W/cm$^2$ 
}
 \label{fig:current_kshifted2_p1d10}
\end{figure}

Figure \ref{fig:current_kshifted2_p1d10} (a) and (b) show
currents and excitation energies in the unit cell using 
the $k$-shifted basis expansion.
For comparison, results of the 3D grid calculation and 
of the $k$-fixed basis function with the largest number of unoccupied 
orbitals are shown.

As in the linear response calculation shown in the previous section, 
we employ occupied orbitals of shifted crystalline momenta, 
$e \Delta A/ \hbar c = \pm 0.01$. 
This value is comparable to the maximum value of the vector
potential in Eq. (\ref{eq:laser}) in the present calculation, 
$A(t)/ \hbar c=0.0047$ a.u., at $t=5.33$ fs. 

As seen from both panels (a) and (b) of 
Fig. \ref{fig:current_kshifted2_p1d10}, 
calculations using $k$-shifted basis expansion well reproduce 
results of the 3D grid calculation.
Although much more orbitals are used, results 
in the $k$-fixed basis expansion are qualitatively wrong.
They are consistent with those in the linear response
calculations where, as shown in Fig. \ref{fig:eps_Re_kshifted}, 
the real part of the dielectric function at $\hbar\omega=3.10$ eV 
is well described in the $k$-shifted basis expansion.

\subsection{Electron dynamics under a strong pulsed electric field}

We next move to the case of electron dynamics in $\alpha$-quartz under
a strong pulsed electric field. 
The maximum intensity of the pulsed electric field is 
set to $5\times 10^{12}$ W/cm$^2$.
For this electric field, the maximum absolute value of the vector 
potential of Eq. (\ref{eq:laser}) reaches $eA(t)/\hbar c= 0.1$ a.u. at $t=5.33$ fs.

We examine two kinds of $k$-shifted basis expansions.
One is the same as those employed in the linear response 
calculations of previous subsection, $k$-shifted-2 $(N_{unocc})$, in which
two shifted $k$-points of $e\Delta A/\hbar c =(0,0,\pm 0.01)$
are used.
The other is $k$-shifted-4 $(N_{unocc})$ in which four shifted $k$-points
of $e\Delta A/\hbar c = (0,0, \pm 0.01)$ and $(0,0, \pm 0.15)$ are 
used. We note that the shift of the crystalline momentum 0.15 a.u.
is similar in magnitude to the maximum value of the vector potential, 
0.1 a.u.

\begin{figure}[tb]
 \begin{center}
  \includegraphics[width= 8cm]{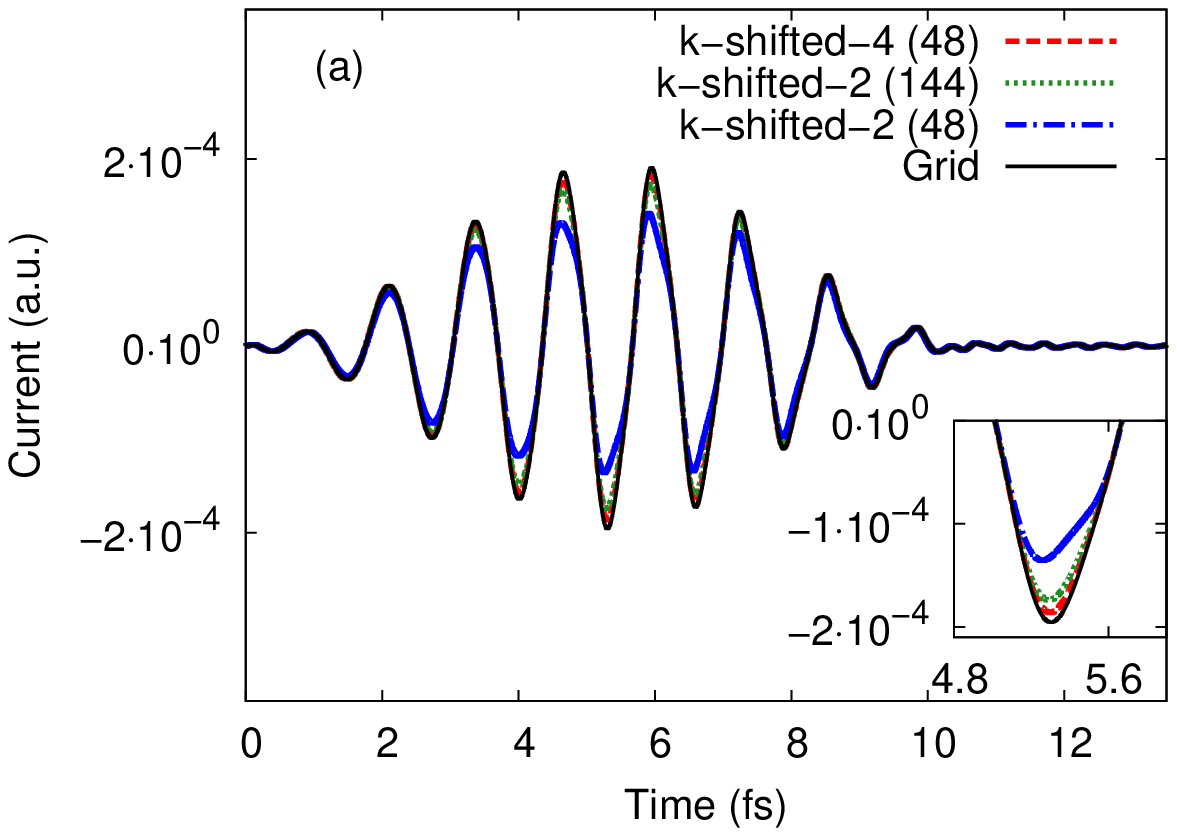}
  \includegraphics[width= 8cm]{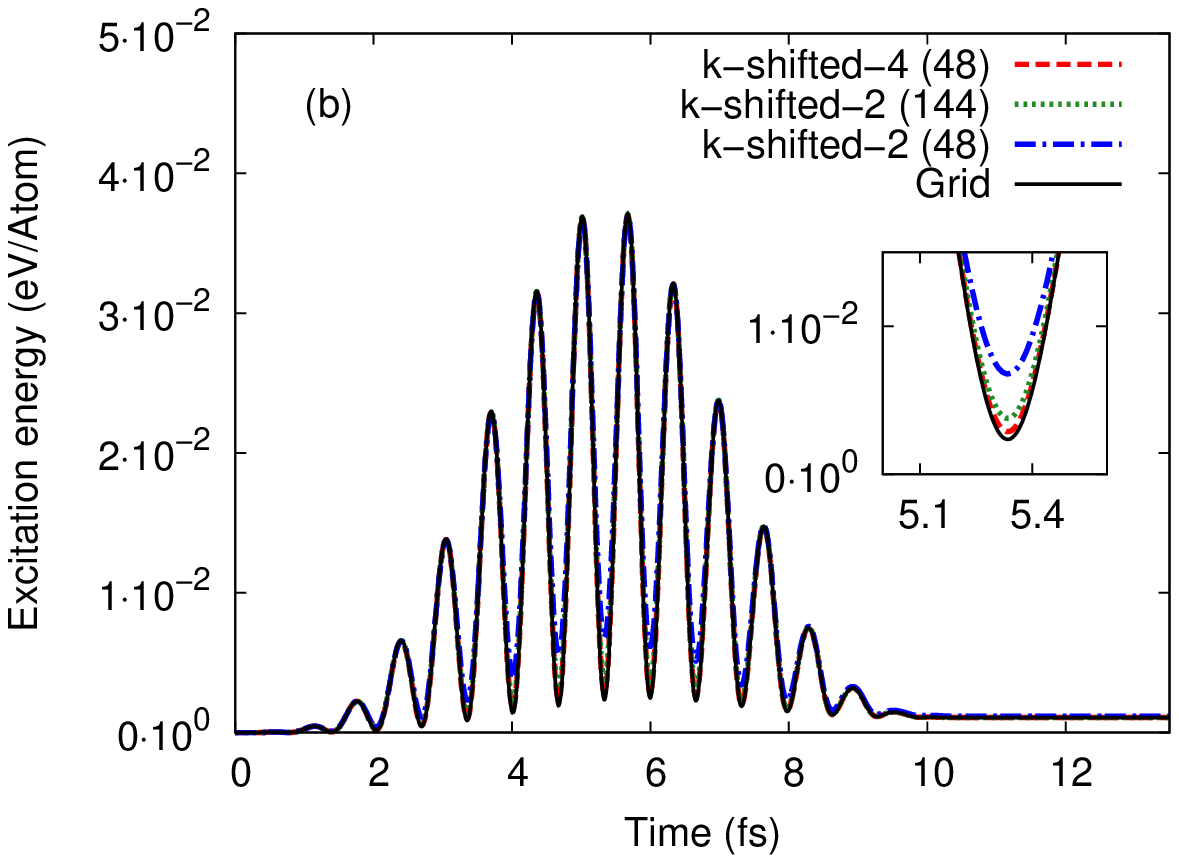}
 \end{center}

 \caption{ Currents (a) and excitation energies per unit cell (b)
as functions of time are shown.
Black-solid line shows results of the 3D grid calculation,
while other lines show results of the $k$-shifted basis expansions.
The maximum intensity of the pulsed electric field is set to
$I=5.0\times 10^{12}$ W/cm$^2$.
}
 \label{fig:current_kshifted4_p5d12}
\end{figure}

Figure \ref{fig:current_kshifted4_p5d12} shows electric currents and 
excitation energies calculated using $k$-shifted-2 (48) basis 
expansion, $k$-shifted-2 (144) basis expansion, and 
$k$-shifted-4 (48) basis expansion. 
For comparison, results in the 3D grid calculation are also shown.
We note that the total number of the basis functions is 192
for $k$-shifted-2 (48), and 288 for both $k$-shifted-2 (144)
and $k$-shifted-4 (48).
At first glance, three calculations of $k$-shifted-2 (48),
$k$-shifted-2 (144), and $k$-shifted-4 (48) work reasonably.
However, a closer look indicates that the $k$-shifted-2 (48)
basis expansion, which was successful for the weak pulse case,
is not accurate enough.
The maximum value of the current is not well reproduced, as seen
from (a).
In contrast, the $k$-shifted-4 (48) and the $k$-shifted-2 (144)
basis expansions well reproduce the results of the 3D grid calculation.
The $k$-shifted-4 (48) basis expansion provides slightly better 
description compared with the $k$-shifted-2 (144) basis expansion.

We show one more case using a pulsed electric field whose frequency is lower 
than, but the maximum value of the vector potential is the same as 
that used in Fig. \ref{fig:current_kshifted4_p5d12}.
The maximum intensity and the mean frequency of the pulsed electric
field are set to
$1.25 \times 10^{12}$ W/cm$^2$ and $1.55$ eV/$\hbar$, respectively.

\begin{figure}[tb]
 \begin{center}
  \includegraphics[width= 8cm]{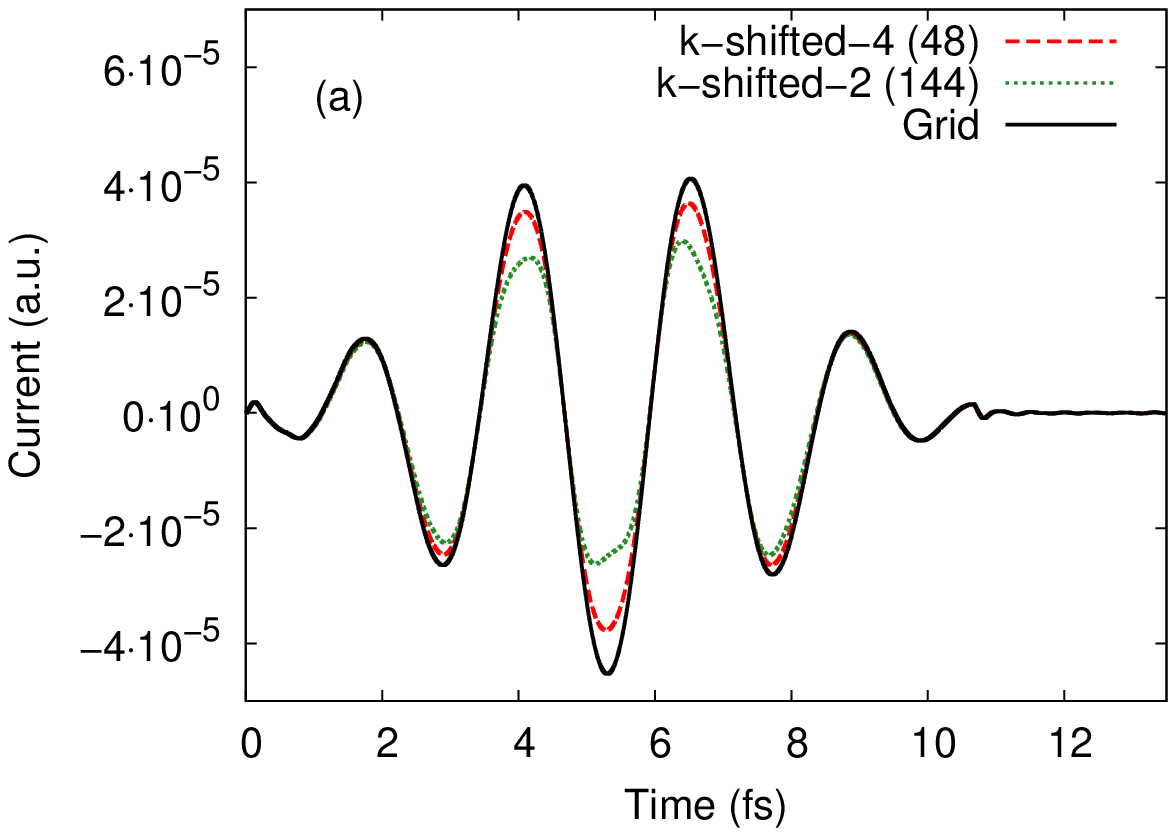}
  \includegraphics[width= 8cm]{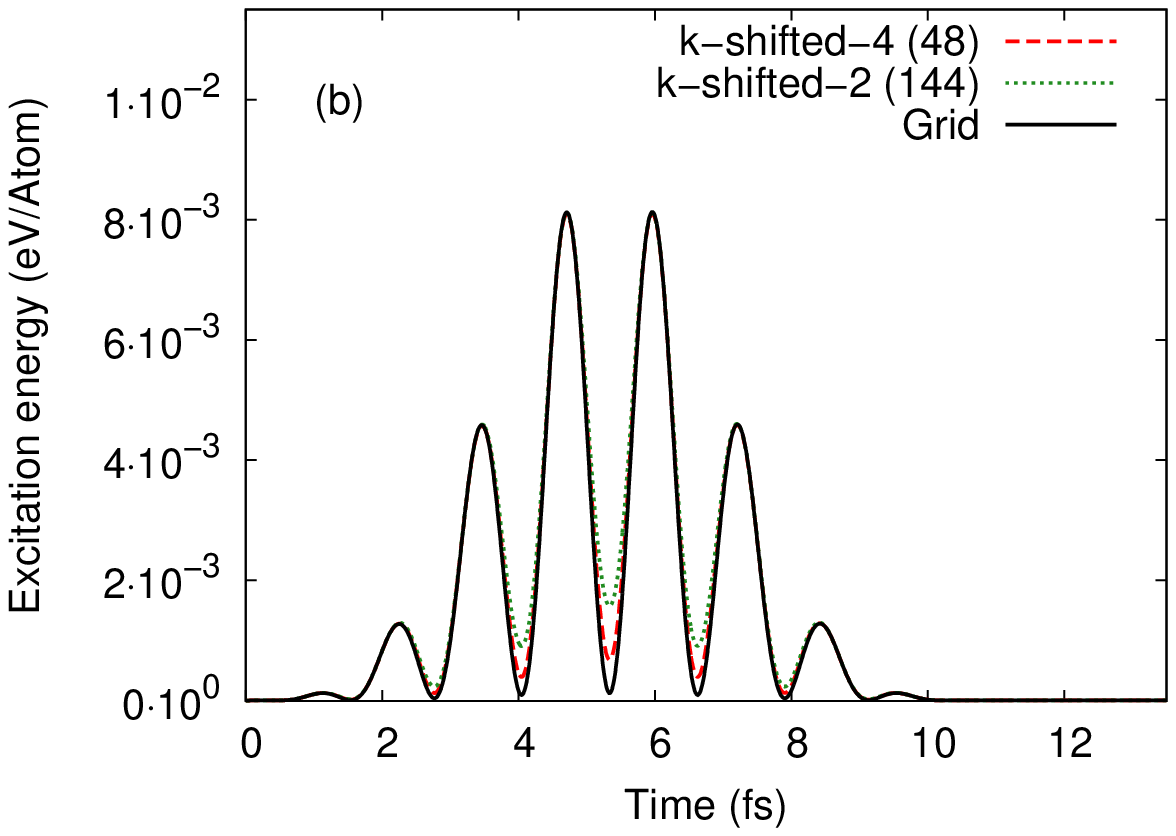}
 \end{center}

 \caption{
Currents (a) and excitation energies per unit cell (b) as functions
of time are shown.
Black-solid line shows results of the 3D grid calculation,
red-dashed line shows results of the $k$-shifted-4 (48) basis expansion, 
and green-dotted line shows results of the $k$-shifted-2 (144) basis
expansion. The maximum intensity and the mean frequency of the pulsed
electric field is set to  $I=1.25 \times 10^{12}$W/cm$^2$ and 
$\omega = 1.55$eV/$\hbar$, respectively.}
 \label{fig:current_kshifted4_p125d12_155ev}
\end{figure}

Figure \ref{fig:current_kshifted4_p125d12_155ev} shows the currents 
and the excitation energies as functions of time using basis
functions of $k$-shifted-4 (48) and $k$-shifted-2 (144).
The figure shows that results of the $k$-shifted-4 (48) basis expansion 
show better agreement with the 3D grid calculation than those of the 
$k$-shifted-2 (144) basis expansion.
This fact indicates that basis functions which include occupied orbitals 
of more $k$-shifted points are required for electron dynamics induced 
by pulsed electric field of lower frequencies.

\section{Computational aspects}
In this section, we discuss computational speed up of 
the electron dynamics calculation using the basis expansion method.
Here we compare the computational time of $k$-shifted-4 (48) basis 
expansion with that of the 3D grid representation in $\alpha$-quartz
calculations. 
We note that the number of basis functions and the number of 
grid points required to describe accurately the electron dynamics 
depend on the material and on the choice of the parameters
of pulsed electric fields. 

There are two aspects in the basis expansion calculations 
that contribute to speed up the electron dynamics calculations.
One is the reduction of the matrix dimension.
In 3D grid calculations, the operation of the Hamiltonian
on orbitals is an operation of a large-sparse matrix on a vector,
while in the $k$-shifted basis expansion, a dense matrix operates
on a vector.
In the present $\alpha$-quartz case, the 3D grid calculations employ 
$36,000=20 \times 36 \times 50$ grid points to describe the Bloch
orbitals. This means that the Hamiltonian is a $36,000\times 36,000$  
sparse matrix. This is reduced to a $288 \times 288$ dense matrix 
in the $k$-shifted-4 (48) basis expansion.
The other is a possible use of longer stride of time in the real-time
evolution. In explicit methods such as the Taylor expansion method
which we usually adopt in the 3D grid representation, the stride 
which provides stable time evolution is limited by the maximum
eigenvalue of the Hamiltonian. In the 3D grid representation, 
the maximum eigenvalue may be estimated from the grid spacing
$\Delta x$ by $E_{max} = 3  \hbar^2 (\pi/\Delta x)^2/2m$.
The maximum eigenvalue in the orbital expansion methods is much 
smaller than that of 3D grid representation.

We first investigate the speed up coming from the reduction of 
the matrix dimension. We compare the computational time of the real-time
evolution, employing the same stride of time step, $\Delta t=0.02$ a.u. 
and the number of time steps, $N_t=28,000$. This time step is appropriate 
in the 3D grid representation. We confirmed that the time evolution 
calculation is not stable if we use a larger step, $\Delta t=0.03$ a.u., 
in the 3D grid representation. The time evolution calculation is
carried out at the T2K-Tsukuba supercomputer, University of Tsukuba. 
We employed 64 CPU cores to parallelize the calculation distributing
different $k$-points, $64 = 4\times 4\times 4$ $k$-points, into cores.
We have found that the calculation in the 3D grid representation costs 
16.71 hours, while the calculation in the $k$-shifted-4 (48) basis 
expansion costs 1.05 hours. Therefore the speed up by reducing the 
matrix dimension is about 15 times.

We note that the reduction of the computational time is not simply 
related to the dimension of the matrix. 
Since the matrix is sparse in the 3D grid representation while 
it is dense in the basis expansions, the computational time is 
much smaller for the sparse matrix if the matrix dimensions are the same.

We next investigate the possibility of employing longer stride
for the time step. For the $k$-shifted-4 (48) basis expansion calculation,
we have found that we may use $\Delta t = 0.07$ a.u., while the
calculation employing $\Delta t= 0.08$ a.u. results in divergence
after a short evolution. Therefore, we can use about 3.5 times
longer stride of time in the $k$-shifted basis expansion than that
in the 3D grid representation. Combining the two speed up factors, 
$15\times 3.5$, we can speed up about 50 times in the $k$-shifted 
basis expansion compared with the 3D grid calculation.

\section{Summary and conclusion}

In this paper, we explored efficient basis expansion methods to 
describe electron dynamics in crystalline solids under an intense 
and ultrashort laser pulse irradiation as well as under a weak
field in the linear response. We utilize a realistic Hamiltonian 
constructed by the first-principles density functional theory. 
Employing uniform grid representation in the three-dimensional Cartesian 
coordinates, we can calculate the evolution of orbitals robustly. 
However, it costs excessive computational costs.

One may expect that a simple and natural choice for the basis functions 
would be eigenfunctions of static Kohn-Sham Hamiltonian. However, it 
turned out that the convergence is very slow with respect to the number
of orbitals employed. 
We then propose to add occupied orbitals of nearby $k$-points
to the occupied and unoccupied orbitals at original $k$-points 
to mimic the use of Houston functions.
We call this procedure the $k$-shifted basis expansion.
It was found that the $k$-shifted basis expansion dramatically improves
the description. 
We demonstrate usefulness of the $k$-shifted basis expansion numerically, 
taking examples of responses of SiO$_2$ under weak and strong pulsed 
electric fields. We also clarify analytically the reason why the use of 
the $k$-shifted occupied orbitals improves the description. 

We also investigated the computational speed up using the $k$-shifted 
basis expansion. For the electron dynamics calculation of SiO$_2$, 
we found a speed up of about 50 times is achieved employing the $k$-shifted basis 
functions, in comparison with the 3D grid representation.
We expect this method will be much beneficial in the calculations
of laser-matter interactions, especially when we couple the dynamics of 
macroscopic electronmagnetism and microscopic electron dynamics.

\section*{Acknowledgments}
This work is supported by the Grants-in-Aid for Scientific Research Nos. 23340113 and 25104702.
The numerical calculations were performed on the supercomputer at the Institute of Solid State Physics, 
University of Tokyo, and T2K-Tsukuba at the Center for Computational Sciences, University of Tsukuba.


\end{document}